\documentstyle[12pt]{article}
\newcommand{\bea}{\begin{eqnarray}}
\newcommand{\eea}{\end{eqnarray}}
\newcommand{\be}{\begin{equation}}
\newcommand{\ee}{\end{equation}}

\newcommand{\nn}{\nonumber}
\newcommand{\rf}[1]{(\ref{#1})}

\begin{document}

\begin{center}
{\large CONFIRMATOIN OF THE $\sigma$-MESON BELOW 1 GEV AND INDICATION FOR
THE $f_0(1500)$ GLUEBALL} \\
\bigskip

{\large Yu.S.~Surovtsev}\\
\bigskip
Bogoliubov Laboratory of Theoretical Physics, Joint Institute for Nuclear
Research, Dubna 141 980, Moscow Region, Russia\\
\bigskip
{\large D.~Krupa} and {\large M.~Nagy}\\
\bigskip
Institute of Physics, Slov.Acad.Sci., D\'ubravsk\'a cesta 9,
842 28 Bratislava, Slovakia
\end{center}
\begin{quote}
\baselineskip 10pt
On the basis of a simultaneous description of the isoscalar $s$-wave channel
of the $\pi\pi$ scattering (from the threshold up to 1.9 GeV) and of the
$\pi\pi\to K\overline{K}$ process (from the threshold to $\sim$ 1.4 GeV) in
the model-independent approach, a confirmation of the $\sigma$-meson at
$\sim$ 660 MeV and an indication for the glueball nature of the $f_0(1500)$
state are obtained.
\end{quote}

A problem of scalar mesons is most troublesome and long-lived in the light
meson spectroscopy. Among difficulties in understanding the scalar-isoscalar
sector there is the one related to a strong model-dependence of information on
multichannel states obtained in analyses based on the specific dynamic models
or using an insufficiently-flexible representation of states ({\it e.g.}, the
standard Breit -- Wigner form). Earlier, we have shown \cite{KMS-nc96} that an
inadequate description of multichannel states gives not only their distorted
parameters when analyzing data but also can cause the fictitious states when
one neglects important (even energetic-closed) channels. In this paper we are
going, conversely, to demostrate that the large background ({\it e.g.}, that
happens in analyzing $\pi\pi$ scattering), can hide low-lying states, even
such important for theory as a $\sigma$-meson \cite{PDG-98}. The latter is
required by most of the models (like the linear $\sigma$-models or the
Nambu -- Jona-Lasinio models \cite{NJL}) for spontaneous breaking of chiral
symmetry.
Since earlier all the analyses of the $s$-wave $\pi\pi$ scattering gave a
large $\pi\pi$-background, it was said that this state (if exists) is
"unobservably"-wide. Recently, new analyses of the old and new experimental
data have been performed which give a very wide scalar-isoscalar state in the
energy region 500-850 MeV \cite{Zou}. However, these analyses
use either the Breit -- Wigner form (even if modified) or specific forms of
interactions in a quark model, or in a multichannel approach to the considered
processes; therefore, there one cannot talk about a model independence of
results. Besides, in these analyses, a large $\pi\pi$-background is obtained.
We are going to show that a proper detailing of the background (as allowance
for the left-hand branch-point) permits us to extract from the latter a very
wide (but observable) state below 1 GeV.

An adequate consideration of multichannel states and a model-independent
information on them can be obtained on the basis of the first principles
(analyticity, unitarity and Lorentz invariance) immediately applied to
analyzing experimental data. The way of realization is a consistent allowance
for the nearest singularities on all sheets of the Riemann surface of the
$S$-matrix. The Riemann-surface structure is taken into account by a proper
choice of the uniformizing variable. Earlier, we have proposed this method for
2- and 3-channel resonances and developed the concept of standard clusters
(poles on the Riemann surface) as a qualitative characteristic of a state
and a sufficient condition of its existence as well as a criterion of a
quantitative description of the coupled-process amplitudes when all the
complifications of the analytic structure due to a finite width of resonances
and crossing channels and high-energy ``tails'' are accumulated in quite a
smooth background \cite{KMS-nc96}.
Let us stress that for a wide state, the pole position (the pole cluster one
for multichannel states) is a more stable characteristic than the mass and
width which are strongly dependent on a model. The cluster kind is determined
from the analysis of experimental data and is related to the state nature. At
all events, we can,
in a model-independent manner, discriminate between bound states of particles
and the ones of quarks and gluons, qualitatively predetermine the relative
strength of coupling of a state with the considered channels, and obtain an
indication on its gluonium nature.

In this work, we restrict ourselves to a two-channel approach when considering
simultaneously the coupled processes $\pi\pi\to \pi\pi,K\overline{K}$.
Therefore, we have the two-channel $S$-matrix determined on the 4-sheeted
Riemann surface. The $S$-matrix elements $S_{\alpha\beta}$, where
$\alpha,\beta=1(\pi\pi), 2(K\overline{K})$, have the right-hand (unitary) cuts
along the real axis of the $s$-variable complex plane, starting at the points
$4m_\pi^2$ and $4m_K^2$ and extending to $\infty$, and the left-hand cuts,
which are related to the crossing-channel contributions and extend along the
real axis towards $-\infty$ and begin at $s=0$ for $S_{11}$ and at
$4(m_K^2-m_\pi^2)$ for $S_{22}$ and $S_{12}$. We number the Riemann-surface
sheets according to the signs of analytic continuations of the channel
momenta ~$k_1=(s/4-m_\pi^2)^{1/2}, ~k_2=(s/4-m_K^2)^{1/2}$~ as follows: ~
signs $({\mbox{Im}}k_1,{\mbox{Im}}k_2)=++,-+,--,+-$ correspond to the sheets
I,II,III,IV.

To elucidate the resonance representation on the Riemann surface, we express
analytic continuations of the matrix elements to the unphysical sheets
$S_{\alpha\beta}^L$ ($L=II,III,IV$) in terms of them on the physical sheet
$S_{\alpha\beta}^I$. Those expressions are convenient for our purpose because,
on sheet I (the physical sheet), the matrix elements $S_{\alpha\beta}^I$ can
have only zeros beyond the real axis. Using the reality property of the
analytic functions and the 2-channel unitarity, one can obtain
\bea \label{S_L}
&&S_{11}^{II}=\frac{1}{S_{11}^I},\qquad ~~~~S_{11}^{III}=\frac{S_{22}^I}{\det S^I},
\qquad S_{11}^{IV}=\frac{\det S^I}{S_{22}^I},\nn\\
&&S_{22}^{II}=\frac{\det S^I}{S_{11}^I},\qquad S_{22}^{III}=\frac{S_{11}^I}
{\det S^I},\qquad S_{22}^{IV}=\frac{1}{S_{22}^I},\\
&&S_{12}^{II}=\frac{iS_{12}^I}{S_{11}^I},\qquad ~~~S_{12}^{III}=\frac{-S_{12}^I}
{\det S^I},\qquad S_{12}^{IV}=\frac{iS_{12}^I}{S_{22}^I},\nn
\eea
Here $\det S^I=S_{11}^I S_{22}^I-(S_{12}^I)^2$.
In the matrix element, a resonance with the only decay mode is represented by a
pair of complex-conjugate poles on sheet II as the nearest singularities and a
pair of conjugate zeros on sheet I at the same points of complex energy. In
the 2-channel case, the mentioned formulae of analytical continuations
immediately give the resonance representation by poles and zeros on the
4-sheeted Riemann surface. One must discriminate between three types of
resonances described by a pair of conjugate zeros on sheet I: ({\bf a}) in
$S_{11}$, ({\bf b}) in $S_{22}$, ({\bf c}) in each of $S_{11}$ and $S_{22}$.
A resonance of every type is represented by a pair of complex-conjugate
clusters (of poles and zeros on the Riemann surface) of size typical of
strong interactions. Thus, we arrive at the notion of three standard
pole-clusters which represent two-channel bound states of quarks and gluons.
Note that this resonance division into types is not formal. In paricular,
the resonance, coupled strongly with the first ($\pi\pi$) channel, is
described by the pole cluster of type ({\bf a}); if the resonance is coupled
strongly with the $K\overline{K}$ and weakly with $\pi\pi$ channel (say, if it
has a dominant $s\overline{s}$ component), then it is represented by the
cluster of type ({\bf b}); finally, since a most noticeable property of a
glueball is the flavour-singlet structure of its wave function and, therefore,
(except the factor $\sqrt{2}$ for a channel with neutral particles)
practically equal coupling with all the members of the nonet, then a glueball
must be represented by the pole cluster of type ({\bf c}) as a necessary
condition.

Just as in the 1-channel case, the existence of a particle bound-state means
the presence of a pole on the real axis under the threshold on the physical
sheet, so in the 2-channel case, the existence of a bound state in channel 2
($K\overline{K}$ molecule), which, however, can decay into channel 1
($\pi\pi$ decay), would imply the presence of a pair of complex conjugate
poles on sheet II under the threshold of the second channel without an
accompaniment of the corresponding shifted pair of poles on sheet III
\cite{MP-92}.

In our previous 2-channel analysis of the $\pi\pi$ scattering, we have
obtained satisfactory description ($\chi^2/\mbox{ndf}\approx1.00$) with two
resonances ($f_0 (975)$ and $f_0 (1500)$) and with the large
$\pi\pi$-background. There, in the uniformizing variable, we have taken into
account only the right-hand branch-points at $s=4m_\pi^2$ and $s=4m_K^2$.

Now, to take also the left-hand branch-point at $s=0$ into account, we use
the uniformizing variable
\be \label{v}
v=\frac{m_K\sqrt{s-4m_\pi^2}+m_\pi\sqrt{s-4m_K^2}}{\sqrt{s(m_K^2-m_\pi^2)}},
\ee
which maps the 4-sheeted Riemann surface onto the $v$-plane, divided into two
parts by a unit circle centered at the origin. The sheets I (II), III (IV)
are mapped onto the exterior (interior) of the unit disk on the upper and
lower $v$-half-plane, respectively. The physical region extends from the point
$i$ on the imaginary axis ($\pi\pi$ threshold) along the unit circle clockwise
in the 1st quadrant to point 1 on the real axis ($K\overline{K}$ threshold)
and then along the real axis to point $b=\sqrt{(m_K+m_\pi)/(m_K-m_\pi)}$ into
which $s=\infty$ is mapped on the $v$-plane. The intervals
$(-\infty,-b],[-b^{-1},b^{-1}],[b,\infty)$ on the real axis are the images of
the corresponding edges of the left-hand cut of the $\pi\pi$-scattering
amplitude. The type ({\bf a}) resonance is represented in $S_{11}$ by two
pairs of the poles on the images of the sheets II and III, symmetric to each
other with respect to the imaginary axis, by zeros, symmetric to these poles
with respect to the unit circle.

Note that the variable $v$ is uniformizing for the $\pi\pi$-scattering
amplitude, however, the amplitudes of the $K\overline{K}$ scattering and
$\pi\pi\to K\overline{K}$ process do have the cuts on the $v$-plane, which
arise from the left-hand cut on the $s$-plane, starting at
$s=4(m_K^2-m_\pi^2)$. Under conformal mapping \rf{v}, this left-hand cut is
mapped into cuts which begin at the points
$v=(m_K\sqrt{m_K^2-2m_\pi^2}\pm im_\pi)/(m_K^2-m_\pi^2)$
on the unit circle on the $v$-plane, go along it up to the imaginary axis, and
occupy the latter. This left-hand cut will be neglected in the Riemann-surface
structure, and the contribution on the cut will be taken into account in the
$K\overline{K}$ background as a pole on the real $s$-axis on the physical
sheet in the sub-$K\overline{K}$-threshold region; on the $v$-plane, this pole
gives two poles on the unit circle in the upper half-plane, symmetric to each other
with respect to the imaginary axis, and two zeros, symmetric to the poles with
respect to the real axis, {\it i.e.} at describing the process $\pi\pi\to
K\overline{K}$, one additional parameter is introduced, say, a position $p$ of
the zero on the unit circle.

For the simultaneous analysis of experimental data on the coupled processes it
is convenient to use the Le Couteur-Newton relations \cite{LN} expressing the
$S$-matrix elements of all coupled processes in terms of the Jost matrix
determinant $d(k_1,k_2)\equiv d(s)$, the real analytic function with the only
square-root branch-points at the process thresholds $k_i=0$.

On $v$-plane the Le Couteur-Newton relations are
\be \label{v:C-Newton}
S_{11}=\frac{d(-v^{-1})}{d(v)},\quad S_{22}=\frac{d(v^{-1})}{d(v)},
\quad S_{11}S_{22}-S_{12}^2=\frac{d(-v)}{d(v)}
\ee
with the $d$-function that on the $v$-plane already does not possess
branch-points is taken as
\be \label{d}
d=d_B d_{res},
\ee
where ~$d_B=B_{\pi}B_K$; $B_{\pi}$ contains the possible remaining
$\pi\pi$-background contribution, related to exchanges in crossing channels;
$B_K$ is that part of the $K\overline{K}$ background which does not contribute
to the $\pi\pi$-scattering amplitude
\be \label{B_K}
B_K=v^{-1}(1-pv)(1+p^*v).
\ee
The function $d_{res}(v)$ represents the contribution of resonances, described
by one of three types of the pole-zero clusters, {\it i.e.}, except for the
point $v=0$, it consists of zeros of clusters:
\be \label{d_res}
d_{res} = v^{-M}\prod_{n=1}^{M} (1-v_n^* v)(1+v_n v),
\ee
where
$n$ runs over the independent zeros; therefore, for resonances of the types
({\bf a}) and ({\bf b}), $n$ has two values, for the type ({\bf c}), four
values; $M$ is the number of pairs of the conjugate zeros.

On the basis of these formulas, we analyze simultaneously the available
experimental data on the $\pi\pi$-scattering \cite{Hyams,edps} and
the process $\pi\pi\to K\overline{K}$ \cite{Wickl} in the channel with
$I^GJ^{PC}=0^+0^{++}$.

To obtain the satisfactory description ($\chi^2/\mbox{ndf}\approx 2.2$) of the
$s$-wave $\pi\pi$ scattering from the threshold to 1.89 GeV and $|S_{12}|$
from the threshold to 1.4 GeV (where the 2-channel unitarity is valid), we
have taken $B_\pi=1$ in eq.\rf{d}, and three multichannel resonances turned
out to be sufficient:  the two ones of the type ({\bf a}) ($f_0 (660)$ and
$f_0 (980)$) and $f_0 (1500)$ of the type ({\bf c}).  A satisfactory
description of the phase shift of the $\pi\pi\to K\overline{K}$ matrix element
is obtained to 1.5 GeV with the value of the parameter $p=0.993613-0.112842i$
(this corresponds to the position of the pole on the $s$-plane at
$s=4(m_K^2-m_\pi^2)-0.06$).

In the table, the obtained parameter values of poles on the corresponding
sheets of the Riemann surface are cited on the complex energy plane
($\sqrt{s_r}= {\rm E}_r-i\Gamma_r /2$). We stress that these are not masses
and widths of resonances. Since, for wide resonances, values of masses and
widths are very model-dependent, it is reasonable to report characteristics of
pole clusters which must be rather stable for various models.  %
\begin{table}[htb]
\begin{center}
\begin{tabular}{|c|rl|rl|rl|rl|rl|rl|}
\hline
{} & \multicolumn{2}{c|}{$f_0 (660)$} & \multicolumn{2}{c|}{$f_0(980)$}
& \multicolumn{2}{c|}{$f_0(1500)$} \\
\cline{2-7}
Sheet & \multicolumn{1}{c}{E, MeV} & \multicolumn{1}{c|}{$\Gamma$, MeV}
& \multicolumn{1}{c}{E, MeV} & \multicolumn{1}{c|}{$\Gamma$, MeV}
& \multicolumn{1}{c}{E, MeV} & \multicolumn{1}{c|}{$\Gamma$, MeV}\\
\hline
II & 600$\pm$14 & 620$\pm$26 & 990$\pm$5 & 25$\pm$10 & 1480$\pm$15
& 400$\pm$35 \\
\hline
III & 720$\pm$15 & 6$\pm$2 & 984$\pm$16 & 200$\pm$32 & 1540$\pm$25
& 300$\pm$35 \\
{} & {} & {} & {} & {} & 1530$\pm$24 & 400$\pm$38 \\
\hline
IV & {} & {} & {} & {} & 1500$\pm$22~ & 300$\pm$34 \\
\hline \end{tabular}
\end{center}
\end{table}
Let us indicate the constant values of the obtained-state couplings with the
$\pi\pi$ and $K\overline{K}$ systems calculated through the residues
of amplitudes at the pole on sheet II.
Taking the resonance part of the amplitude in the form
$T_{ij}^{res}=\sum_r g_{ir}g_{rj}D_r^{-1}(s)$,  where $D_r(s)$ is an
inverse propagator ($D_r(s)\propto s-s_r$), we obtain
(we denote the coupling constants with the $\pi\pi$ and $K\overline{K}$
systems through $g_\pi$ and $g_K$, respectively) for
$f_0(660)$: $g_{\pi}=0.7376\pm 0.12$ GeV and $g_K=0.37\pm 0.1$ GeV, for
$f_0(980)$: $g_{\pi}=0.158 \pm 0.03$ GeV and $g_K=0.86 \pm 0.09$ GeV, for
$f_0(1500)$: $g_{\pi}=0.347 \pm 0.028$ GeV.
In this 2-channel approach, there is no point in calculating the coupling
constant of the $f_0(1500)$ state  with the $K\overline{K}$ system, because
the 2-channel unitarity is valid only to 1.4 GeV, and, above this energy,
there is a considerable disagreement between the calculation of the amplitude
modulus $S_{12}$ and the experimental data.

Let us indicate also scattering lengths calculated in our approach. For the
$K\overline{K}$ scattering, we obtain
~$a_0^0(K\overline{K})=-0.932\pm 0.11+(0.706\pm 0.09)i, ~~m_{\pi^+}^{-1}.$~
A presence of the imaginary part in $a_0^0(K\overline{K})$ reflects the fact,
that already at the threshold of the $K\overline{K}$ scattering, other
channels ($2\pi,4\pi$ etc.) are opened.

For the $\pi\pi$ scattering, we obtain: $a_0^0=0.27\pm 0.06,~~m_{\pi^+}^{-1}.$
Compare with results of some other works both theoretical and experimental:
the value $0.26\pm 0.05$ (L. Rosselet et al.\cite{edps}), obtained in the
analysis of the decay $K\to\pi\pi e\nu$ with using Roy's model;
$0.24\pm 0.09$ (A.A. Bel'kov et al.\cite{edps} from analysis of the
process $\pi^-p\to\pi^+\pi^-n$ with using the effective range formula;
$0.23$ (S. Ishida et al.\cite{Ishida}, modified approach to analysis of
$\pi\pi$ scattering with using Breit-Wigner forms; $0.16$ (S. Weinberg
\cite{Weinberg}, current algebra (non-linear $\sigma$-model)); $0.20$
(J. Gasser, H. Leutwyler \cite{Weinberg}, the theory with the non-linear
realization of chiral symmetry); $0.26$ (M.K. Volkov \cite{Weinberg}, the
theory with the linear realization of chiral symmetry).

So, an existence of the low-lying state with the properties of the
$\sigma$-meson and the obtained value of the $\pi\pi$-scattering length seem
to suggest the linear realization of chiral symmetry.

Here, a parameterless description of the $\pi\pi$ background in the channel
with $I^GJ^{PC}=0^+0^{++}$ is first given.

The $f_0 (1500)$ state is represented by the pole cluster on the Riemann
surface of the $S$-matrix which corresponds to a flavour singlet, {\it e.g.}
the glueball.

Finally, a minimum scenario of the simultaneous description of the processes
$\pi\pi\to\pi\pi,K\overline{K}$ in the channel with $I^GJ^{PC}=0^+0^{++}$ does
not require the ${f_0}(1370)$ resonance; therefore, if this meson exists, it
must be weakly coupled with the $\pi\pi$ channel, {\it e.g.} be the
$s{\bar s}$ state (as to that assignment of the ${f_0}(1370)$ resonance, we
agree with the work \cite{Shakin1}).

This work has been supported by the Grant Program of Plenipotentiary of Slovak
Republic at JINR.  Yu.S. and M.N. were supported in part by the Slovak
Scientific Grant Agency, Grant VEGA No. 2/7175/20; and D.K., by Grant VEGA
No. 2/5085/99.

 \end{document}